\documentclass[12pt]{article}

\newcommand{\msub}[1]{\ensuremath _{\mbox{\scriptsize #1}}} 

\usepackage{graphics}

\begin{document}

\begin{center}{\Large\bf Locality and topology with fat link 
overlap actions}\\[2cm] 
{\em Tam\'as G. Kov\'acs} \\[10mm]
{\em NIC/DESY, Platanenallee 6 \\
D-15738 Zeuthen, Germany}\\
{\sf e-mail: kovacs@ifh.de}\\[4mm]

and \\[4mm]
 
{\em Department of Theoretical Physics\footnote{Address 
                                    after October 1, 2002.} \\ 
     University of P\'ecs \\
     H-7624 P\'ecs, Ifj\'us\'ag \'utja 6. \\[10mm]}
\end{center}

\begin{abstract} 
We study the locality and topological properties of 
fat link clover overlap (FCO) actions.
We find that a small amount of fattening (2-4 steps of APE or
1 step of HYP) already results in greatly improved properties
compared to the Wilson overlap (WO).  We present a detailed study
of the localisation of the FCO and its connection to the 
density of low modes of $A^\dagger A$. In contrast to the
Wilson overlap, on quenched gauge backgrounds we do not find any 
dependence of the localization of the FCO on the gauge coupling.
This suggests that the FCO remains local in
the continuum limit. The FCO also faithfully reproduces the zero mode 
wave functions of typical lattice instantons, not like the Wilson overlap.
After a general discussion of different lattice definitions
of the topological charge we also show that the FCO together with 
the Boulder charge are likely to satisfy the index theorem in the 
continuum limit. Finally, we present a high statistics computation 
of the quenched topological susceptibility with the FCO action.
\end{abstract}

\section{Introduction}

In recent years one of the most important technical and conceptual 
advances in the lattice discretisation of QCD has been
the discovery and practical implementation of chirally
symmetric lattice fermion actions. While in the continuum
limit chiral symmetry is expected to be recovered in
any lattice discretisation of fermions, the new formulation
guarantees that it already holds exactly at non-zero
lattice spacings. This has important practical consequences 
opening new possibilities in the study of QCD with
light quarks. 

By now, there are several different implementations of chirally
symmetric lattice fermions (see e.g.\ Ref.\ \cite{Hernandez:2001yd}),
but being relatively new, the properties of these implementations
have not yet been fully explored. Certainly many questions
are still to be answered before chiral fermions can reach
their full potential. In particular, speed, scaling properties
and locality are just some of the issues that clearly
play an important role in this context and have to be 
optimized before chiral fermions can make their way into
realistic full (unquenched) QCD simulations.

In the present paper we propose to study a simple variant of
one of the formulations of chiral fermions, namely
the overlap \cite{Narayanan:ss}. The basic overlap
construction starts with any ``reasonable'' Dirac operator, $D_0$,
that is used as a ``kernel'' in the overlap construction.
The overlap operator is defined in terms of $D_0$ by the formula
\begin{equation}
 D\msub{ov} = 1 - A \left[A^\dagger A\right]^{-\frac{1}{2}},
 \hspace{1cm} A = 1+s - D_0,
\end{equation}
where $s$ is a real parameter. In the original overlap 
construction $D_0$ was the Wilson lattice Dirac operator.
It is by now clear, however, that properties of the overlap
depend strongly on what operator is used for $D_0$ 
\cite{Hasenfratz:2002rp}-\cite{Bietenholz:2002ks}.
In the construction we propose here, $D_0$ is the clover improved
Dirac operator with the tree-level value of $c\msub{sw}=1.0$ and
smeared fat gauge links \cite{Albanese:ds}. This construction 
is neither particularly imaginative, nor is it new. In particular,
fattened gauge links are used in more complicated 
overlap kernels \cite{Hasenfratz:2002rp,DeGrand:2000tf}
and the type of fat link clover overlap (FCO) we propose here
has already been used to estimate the quark-mass dependence 
of the unquenched topological susceptibility \cite{Kovacs:2001bx}.
The authors of Ref.\ \cite{Kamleh:2001ff} argue that the evaluation of
the FCO---although a slightly different version thereof---is 
roughly two times faster than the Wilson overlap, due to the
better condition number of $A^\dagger A$.

The reason why we propose to study some properties of the FCO
in more detail is that it might represent an optimal compromise
between much more complicated choices of $D_0$ and the Wilson
overlap. In the present paper we shall explicitly demonstrate
that the FCO action has the following advantages compared
to the Wilson Overlap (WO).
   
\begin{itemize}
\item Already after a very small amount of fattening, lattice 
configurations are smooth enough that the parameter $s$
does not have to be tuned, it can be set to zero independently
of the gauge coupling. This feature also makes it possible to
check that the localisation range of  of $D\msub{ov}$ does
not change as the gauge coupling is varied.
\item The operator $A^\dagger A$ is much better conditioned
which translates into savings of a factor of 2--5 in computing
time when evaluating its inverse square root.
\item In contrast to the WO  \cite{Gattringer:2001cf}, the
FCO precisely reproduces the continuum wave function
of quark zero modes in the background of smooth instantons
down to an instanton size of one lattice spacing.
\end{itemize}
   
The downside is, purists might object, that using 
fat gauge links might affect short 
distance properties. This however is not a real problem,
since very little fattening turns out to be enough to ensure
the above listed good properties. We shall also explicitly 
check how non-local the APE smearing is and find that the
needed smearing is actually more localized than the 
overlap itself. Even better, one step of ``HYP'' smearing
\cite{Hasenfratz:2001hp}, a smearing that stays within a
hypercube, is already enough to guarantee the above
listed good properties.

In Section \ref{sec:L} we test the locality of both the
APE smearing and the overlap operator. We find that 
the (exponential) localisation range of 
the FCO is essentially independent 
of the gauge coupling of the background gauge fields in the 
range of (Wilson) $5.7 \leq \beta \leq 6.0$, where we
tested it. The most important consequence of this is
that the FCO action is very likely to remain local
in the continuum limit. In fact, at these gauge couplings
the localisation range of the FCO turns out to be 
somewhat smaller than that obtained for the WO after 
optimizing the parameter $s$.

We also discuss the possible role of small eigenvalues
of $A^\dagger A$ in the locality properties of the overlap.
We find that although the spectral density of $A^\dagger A$
is divergent at zero, small eigenmodes occur with a finite
probability, decreasing towards the continuum limit.
They seem to be related to gauge field
defects living on the scale of the cut-off and the
localisation of the corresponding eigenmodes stays 
constant in lattice units as the continuum limit is
approached. Therefore, they are very unlikely to affect
the locality of the overlap in the continuum limit.

In Section \ref{sec:T} we study the topological properties 
of quenched gauge fields with the FCO operator. We demonstrate
that---unlike the Wilson overlap---the FCO reproduces 
continuum instanton zero mode wave functions down to
lattice instanton sizes of $\rho/a \approx 1-2$.
We also discuss the criteria for a fermionic and a 
gauge field definition of the topological charge 
to satisfy the index theorem in the continuum limit.
We find that the FCO Dirac operator and the Boulder
charge satisfy these criteria. Finally we present a
high statistics computation of the quenched topological 
susceptibility using both definitions of the charge.

\section{Locality}
    \label{sec:L}
Locality is a crucial property of lattice actions since 
it guarantees that physics in the continuum limit is 
independent of the details of the discretisation (universality).
Most of the conventional lattice QCD actions, both
fermionic and gauge actions, are ultralocal. This 
means that they connect only degrees of freedom up to a 
finite distance measured in lattice units. It has
been proved that there is no chirally symmetric 
lattice fermion action that is ultralocal
\cite{Horvath:1998cm}, therefore if we insist on
chiral symmetry, we have to settle with a weaker form
of locality. We demand only that the matrix elements
of the operator should fall off exponentially with the distance
measured in lattice units and that this exponent 
should be non-zero in the continuum limit. 
This weaker form of locality is still enough for
universality. In the present section we study the locality 
of the two ingredients of our FCO action, the smeared gauge
links and the fermion action itself.

\subsection{Smearing}

The smearing that we use on the gauge links entering the
fermion operator consists of two steps. First, each link
is replaced with a linear combination of itself and the
six shortest staples around it, as
\begin{equation}
 U_\mu(x) \hspace{3mm} \longrightarrow \hspace{3mm} U'_\mu(x) = 
    (1-c) U_\mu(x) + \frac{c}{6} \sum_{\nu \neq \mu} 
       \left( S_\mu^{(\nu)}(x) + S_\mu^{(-\nu)}(x) \right),
\end{equation}
where $S_\mu^{(\nu)}(x)$ is the product of gauge links along
the path $(x,x+\hat{\nu},x+\hat{\nu}+\hat{\mu},x+\hat{\mu})$,
the staple and $c$ is a real parameter. Since a linear combination
of SU(3) elements is in general not an element of the group, 
in the second step $U'_\mu(x)$ is projected back to the
gauge group by minimizing tr$U'^\dagger_\mu(x){\cal P}(U'_\mu(x))$
for all possible ${\cal P}(U'_\mu(x)) \in$ SU(3).

One sweep of this procedure through the whole lattice will be
referred to as one step of APE smearing. In the course of a 
sweep the new smeared links are always constructed from the
original ones without immediate replacement. The original gauge links
are replaced with the new smeared ones only after a full sweep
through the lattice has been completed. Therefore, the procedure 
is independent of the order in which the gauge links are 
smeared and it is also strictly ultralocal. 

Since any finite number of APE smearing steps is ultralocal,
from a conceptual point of view operators constructed from smeared 
links are as good as any other 
local operator, as long as the number of smearing steps is kept 
constant in the continuum limit. There is, however, a more
practical issue. In actual simulations, it is desirable to keep
the ``size'' of operators much smaller than the inverse mass of
the heaviest physical particle in the system, otherwise
strong scaling violations can be expected. For this reason,
in the present section we check how the locality of smearing
depends on the number of smearing steps and the smearing 
coefficient, $c$.

The simplest way of doing that is by monitoring how a local
disturbance in the gauge field is propagated into gauge invariant
quantities measured at a distance from the disturbance. We choose
the most local disturbance, a random change of a single 
gauge link and the simplest gauge invariant observable, the
plaquette. After randomly flipping a single gauge link, we 
compute the modulus of the difference between each smeared plaquette of  
the original and the changed configuration. Let us denote this
quantity, averaged over all 24 plaquettes at site $x$, by 
$\delta p(x)$. We are interested in how $\delta p(x)$ 
changes with the distance from the flipped link at $x=0$. 
Therefore, the quantity we look at is $\overline{\delta p}(r)$, 
the average of $\delta p(x)$ over spherical shells of radius $r=|x|$,
normalized such that $\delta p(0)=1$.

\begin{figure}[htb!]
\vspace{3mm}
\begin{center}
\begin{minipage}{140mm}
\resizebox{\textwidth}{!}{
\includegraphics{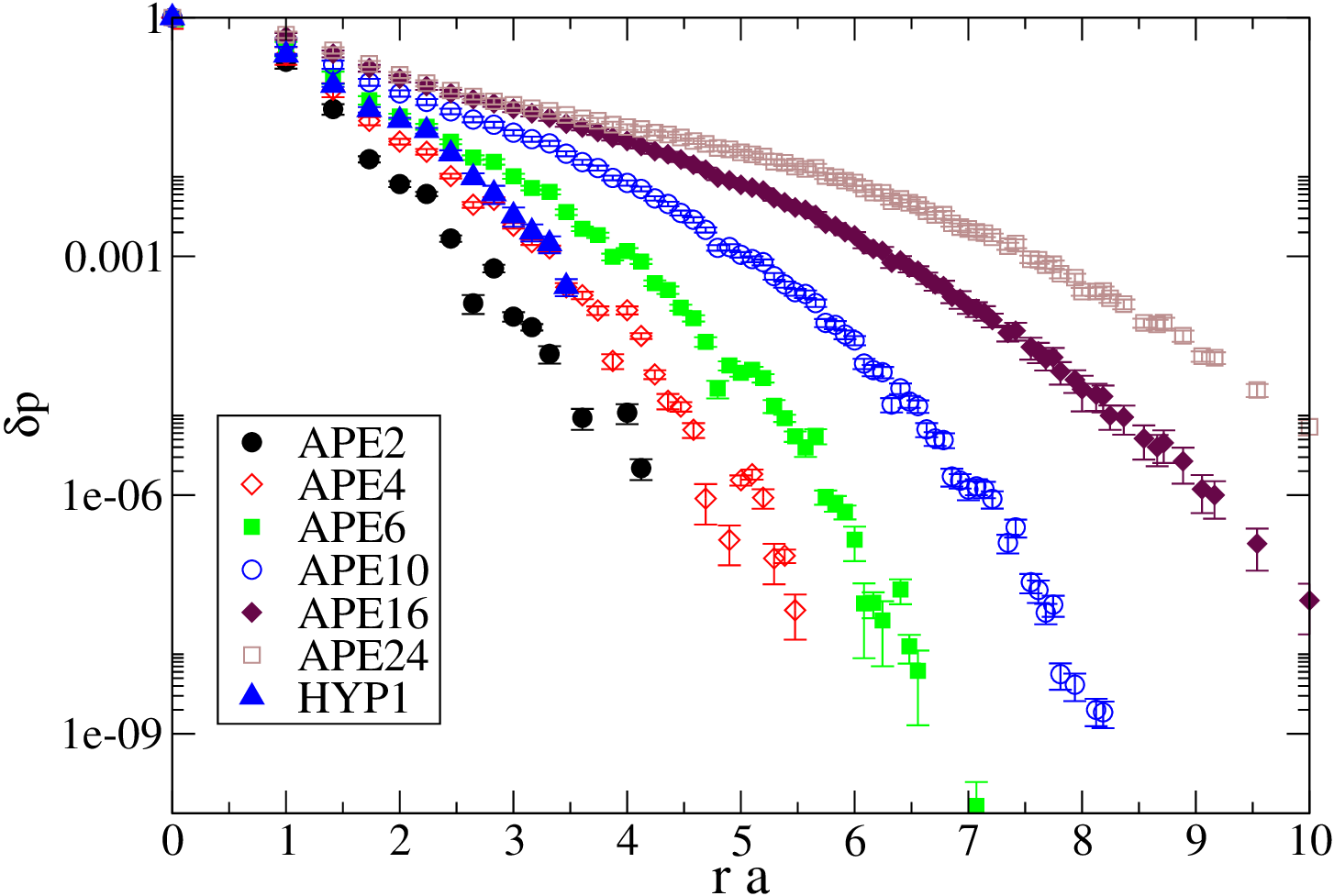}}
\vspace{-0.5cm}\caption{\label{fig:aper} The range of APE smearing
as measured by the change of the average smeared plaquette 
$(\overline{\delta p})$ due to flipping a single link. $ra$ is 
the distance from the flipped link and the curves correspond
to 2,4,6,10,16 and 24 (from left to right) steps of APE smearing
with $c=0.45$, and 1 step of HYP smearing (triangles).}
\end{minipage}
\end{center}
\end{figure}

In Fig.\ \ref{fig:aper} $\overline{\delta p}(r)$ versus $r$
is shown, averaged over a set of 50 quenched Wilson 
$\beta=5.94$ $12^4$ lattice configurations. In this 
computation $c$ was (arbitrarily) set to $0.45$ 
and the number of smearing steps shown are 2,4,6,10,16 and 24.
The important feature we have to notice here is that
for moderate smearing (about $n<10$), $\overline{\delta p}(r)$
decays faster than exponentially over the whole range
where it is non-zero. The starting exponents, that were
fit in the range $1 \leq r \leq 3$ vary from 4.0 to 1.3
for 2 and 10 steps of smearing respectively. Notice that
the actual decay rate of $\overline{\delta p}(r)$ for 
$r\geq 3$ is much faster than that. 

If more than about 10 smearing steps are applied, the function
$\overline{\delta p}(r)$ slowly starts to develop a ``shoulder''
that gradually extends to larger distances. In other words
this means that for more than 10 smearing steps the logarithmic
derivative of $\overline{\delta p}(r)$ is not monotonous; the initial 
short distance exponential decay rate slows down at medium distances 
of $r \approx 3-5$ before accelerating again above that. 

From this behaviour we can conclude that
for most physical quantities, operators built of
2-4-smeared links should be quite safe, and even as much as
10 steps of smearing might not be too severe. Going beyond
this point, however, smearing will start to affect medium
distance properties and might have an adverse effect on spectral
quantities.

For comparison we also plotted $\overline{\delta p}(r)$ for one
step of HYP smearing that is strictly localized within a hypercube
(triangles in Fig.\ \ref{fig:aper}). Regarding the properties
of the smeared configurations, HYP1 is roughly equivalent to
APE4. The initial fall-off of $\overline{\delta p}(r)\msub{HYP1}$ 
is slightly slower than that of APE4, however, 
$\overline{\delta p}(r)\msub{HYP1}$ vanishes for $ra \geq 3.5$. 

\begin{figure}[htb!]
\vspace{3mm}
\begin{center}
\begin{minipage}{110mm}
\resizebox{\textwidth}{!}{
\includegraphics{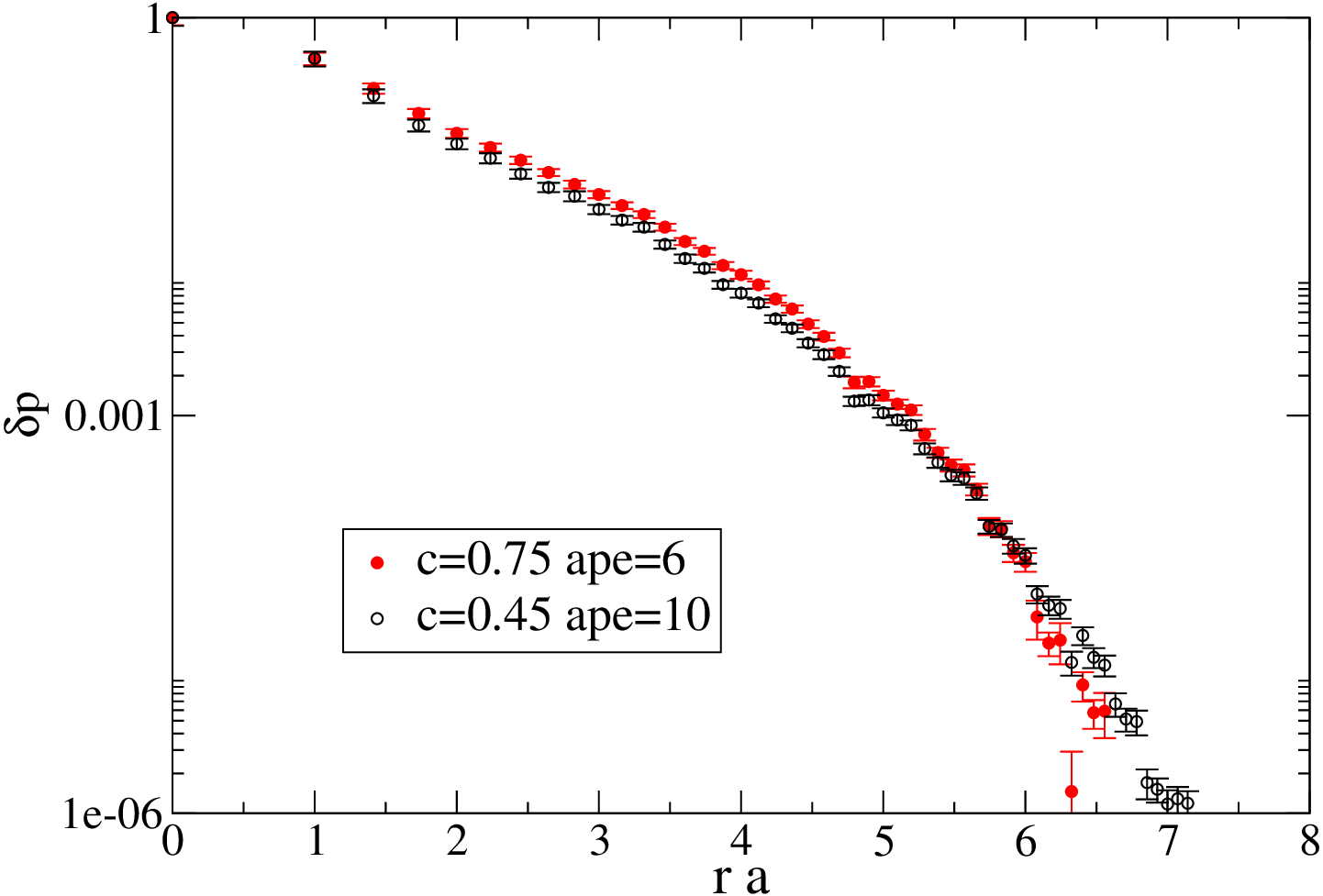}}
\vspace{-0.5cm}\caption{\label{fig:aper_0.75} The same as Fig.\
\ref{fig:aper} but here 6 steps of $c=0.75$ smearing is 
compared to 10 steps of $c=0.45$ APE smearing.}
\end{minipage}
\end{center}
\end{figure}

We also briefly experimented with APE smearing coefficients 
other than $c=0.45$ and confirmed the previously established rule
of thumb that the extent of APE smearing can be adequately 
characterized with $c$ times the number of 
smearing steps\footnote{More precisely, this is true only as
long as the smearing coefficient $c$ is not bigger than its critical
value of $0.75$, which will always be the case here.}. 
To illustrate this point, in Fig.\ \ref{fig:aper_0.75}
we compare the behaviour of $\overline{\delta p}(r)$ obtained with
10 steps of $c=0.45$ and 6 steps of $c=0.75$ APE 
smearing.

\subsection{The fat link clover overlap}

We now turn to testing the locality of the overlap operator
itself. The simplest way to measure how its matrix elements
fall-off with the distance is to create a delta source 
$\psi_k(x) = \delta(x) \delta_{kj}$ ($k$ and $j$ refer to 
Dirac and colour indices) and compute how $||D\msub{ov}\psi (x)||$
decays with the distance from the source.

In the only available detailed study of the locality of the Wilson
overlap, the so called ``taxi driver distance'', the sum of
the moduli of the coordinate differences $r = \sum_\mu |x_\mu|$ was
used and the function
\begin{equation}
  f(r) = \max \left\{ ||D\msub{ov}\psi (x)|| : \sum_\mu|x_\mu|=r \right\}
\end{equation}
was computed \cite{Hernandez:1998et}. To be able to compare our 
results to those of Ref.\ \cite{Hernandez:1998et}, we first also
compute this quantity for the FCO action.

Before presenting our results, however, we have to briefly discuss
how we choose the arbitrary parameter $s$ in the overlap construction.
If the Wilson action is used, the parameter $s$ has to be optimized. 
Roughly speaking, $s$ is an arbitrary ``cut'' that controls whether
a given real eigenvalue of the Wilson operator between the physical
modes and the doublers will become a low physical mode or a doubler
of the resulting overlap. It is clearly advantageous, both from
the practical and the conceptual point of view, to set $s$ in a 
way to minimize the number of small eigenvalues of $A^\dagger A$.
According to Ref.\ \cite{Hernandez:1998et}, e.g.\
at Wilson $\beta=6.0$ the exponent governing the decay of $f(r)$
can vary between $0.28$ and $0.49$ as $s$ changes from $0.0$ to 
$0.4$. In the continuum limit, the optimal value of $s$ is expected 
to go to zero.

Since the localisation range of $D\msub{ov}$ depends strongly on
$s$, it is not an easy task to verify numerically that $D\msub{ov}$ 
remains local in the continuum, i.e. that the exponent goes
to a non-zero constant. To this end, one would need to compute
the optimal exponent at every value of $\beta$ and verify the
statement for that. While the numerical results of Ref.\ 
\cite{Hernandez:1998et} are encouraging in this respect, such
a complete study was not undertaken there. Moreover, it is not
guaranteed that the localization and the density of small 
eigenvalues of $A^\dagger A$ are optimal at the same $s$.

\begin{figure}[htb!]
\vspace{3mm}
\begin{center}
\begin{minipage}{110mm}
\resizebox{\textwidth}{!}{
\includegraphics{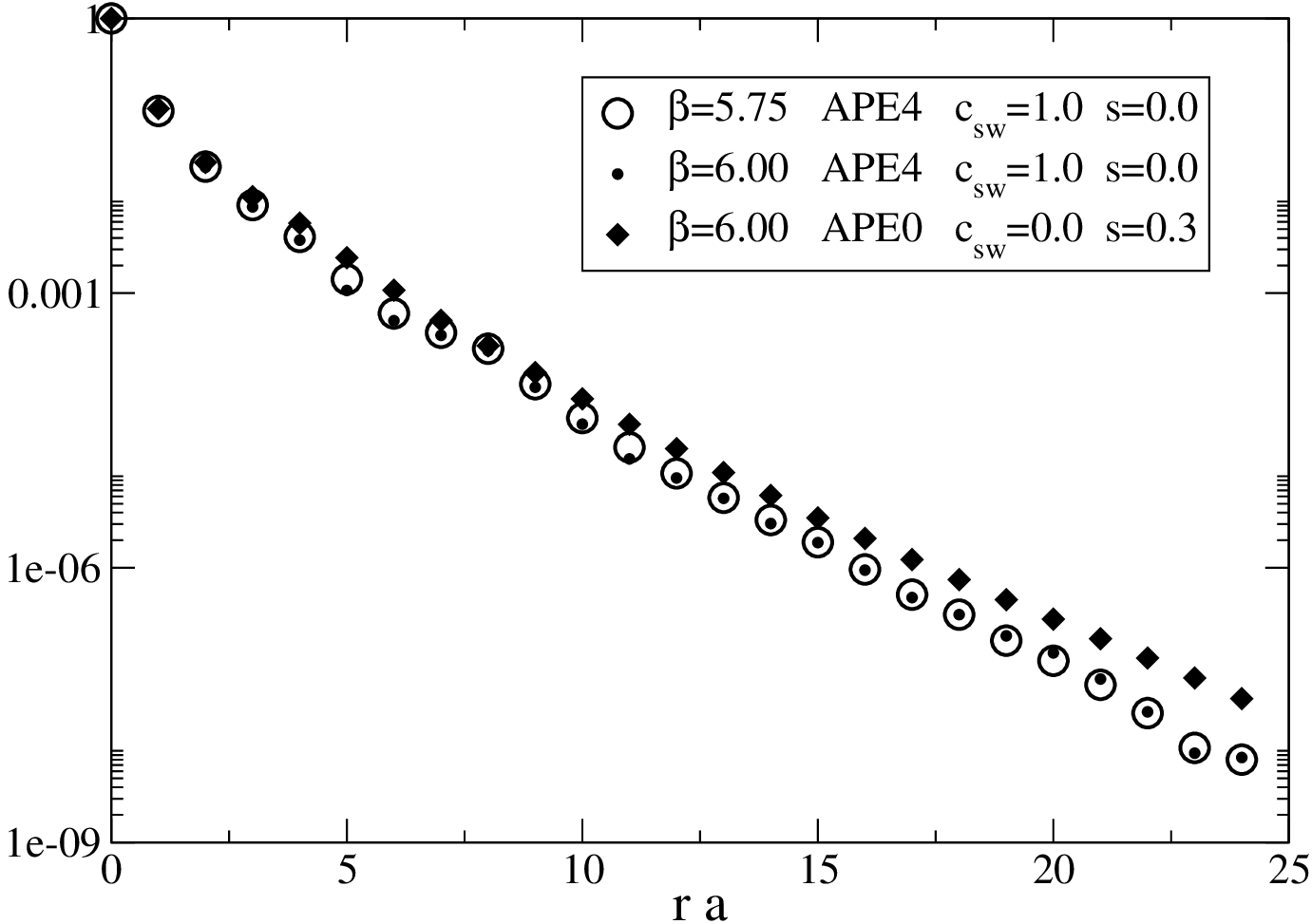}}
\vspace{-0.5cm}\caption{\label{fig:ovrange} $f(r)$, the maximal matrix
element of the overlap as a function of the taxi driver distance $r$.
The small filled and the large empty circles represent the $c\msub{sw}=1$,
$s=0.0$ clover overlap on APE4 gauge backgrounds 
at $\beta=5.75$ and $\beta=6.0$, while the diamonds correspond to 
the Wilson overlap at the optimal value of $s=0.3$ on thin link
$\beta=6.0$ gauge backgrounds.}
\end{minipage}
\end{center}
\end{figure}

The situation with the FCO action is much simpler.
Even after 2-4 steps of APE smearing the gauge field is smooth
enough that $s$ can be safely set to the continuum value, 
zero, regardless of the gauge coupling (at least as long as
the quenched Wilson $\beta \geq 5.70$). In Fig.\ \ref{fig:ovrange}
we compare the function $f(r)$ on three $12^4$ quenched gauge ensembles.
The first two were generated at (Wilson) $\beta=5.75$ 
and one with $\beta=6.00$. The kernel used for the overlap 
here was the clover action
with the perturbative value of $c\msub{sw}=1.0$ and the gauge
links have been 4 times APE smeared with $c=0.45$ (APE4). The two data
sets lie so close to each other that they are hardly 
distinguishable by the naked eye. We also verified that $f(r)$ 
is the same on two other gauge ensembles with intermediate $\beta$
values. This means that a $40\%$ decrease
in the lattice spacing resulted in no discernible change of
the localisation of the overlap; asymptotically it decays 
as $e^{-\nu r}$ with the same exponent, $\nu=0.60$. For comparison
we also included in the same figure the conventional Wilson overlap
on unsmeared $\beta=6.0$ gauge backgrounds. Here the parameter $s$
was chosen to be 0.3 which is roughly the optimal value
for localisation and an exponential fit yielded $\nu=0.51$\footnote{As
in Ref.\ \cite{Hernandez:1998et}, the fit was done for the
range $13 \leq r a \leq 24$ to facilitate the comparison
of the exponents $\nu$. We have to note, however that due
to finite size effects, the obtained values do not coincide
with the asymptotic exponent $\nu$. Nevertheless, they are
accurate enough to assess the qualitative differences between 
various fermion actions.}

Although the FCO appears to be somewhat more local than the
Wilson overlap, the main point here is not this, but 
that in the case of the former, the range of the 
operator (in lattice units) can be seen 
to be independent of the lattice spacing. This strongly suggests
that even on gauge backgrounds generated with the Wilson action
the FCO stays local in the continuum limit.

To appreciate this statement, let us recall what is known about
the locality of the overlap. In Ref.\ \cite{Hernandez:1998et}
it was shown that it is sufficient for its exponential
localisation if the spectrum of $A^\dagger A$ has
a gap at zero. It was also suggested there that a finite gap
might not be necessary, it might be enough if the continuous
spectrum of $A^\dagger A$ is separated from zero. On the other hand,
the spectrum is guaranteed to have a finite gap if the gauge field action
restricts the plaquette values sufficiently close to unity
everywhere. This condition, however, even in the continuum
limit, is not necessarily satisfied for practically useful actions.

Although most practical gauge actions damp plaquette fluctuations
exponentially in the gauge coupling, the 
number of plaquettes in a fixed physical volume also grows 
exponentially in the continuum limit. How the probability of a given
plaquette fluctuation in a fixed physical volume will
behave in the continuum limit, will depend on which of the
two exponentials wins. The results of Ref.\ \cite{Hernandez:1998et}
therefore do not exclude the possibility that the overlap
in gauge backgrounds generated with e.g.\ the Wilson action
becomes non-local in the continuum limit. Even though locality
can be proved for special gauge actions, universality cannot be
used to argue that any other gauge action in the same universality
class will also have the same property. This is because in 
principle, with a different gauge action, the overlap can be 
non-local and universality of the complete fermion gauge system
is not guaranteed then.

\begin{table}[htb!]
\begin{center}
\vspace{3mm}
\begin{tabular}{|l||r|r|r|r|r|} \hline
 $\beta$ & size       & $a$ (fm) & V (fm$^4$) & confs & $\rho(0)$ (fm$^{-4}$)  
                                                            \\ \hline\hline
 5.745 & $8^4$        &  0.153   & 2.27       & 2000  & 1.55(1)      \\ \hline
 5.85  & $10^4, 12^4$ &  0.123   & 2.29 ,4.75 & 3000, 1000 & 0.78(2) \\ \hline
 5.94  & $12^4$       &   0.104  & 2.38       & 1000  & 0.42(2) \\ \hline\hline
\end{tabular}
\end{center}
  \caption{The four gauge field ensembles used for the study of the
spectrum of $A^\dagger A$ and $(A^\dagger A)^{1/2}$. 
The lattice spacing has been set with the Sommer parameter $r_0=0.49$~fm 
and the Wilson gauge couplings were tuned to keep the physical volume 
of the system fixed. The last column lists the spectral density
of $(A^\dagger A)^{1/2}$ at zero. \label{tab:pars}}
\end{table}

Fortunately the situation is much better since neither the
constraint on the plaquette fluctuations nor the one on the spectral
gap of $A^\dagger A$ has been proved to be necessary for locality.
In fact, based on earlier work of the SCRI group \cite{Edwards:1999bm},
the continuous spectrum of $A^\dagger A$ can be expected to
extend all the way down to zero and diverge there as $\lambda^{-1/2}$
at any finite (Wilson gauge action) $\beta$. This is because
in \cite{Edwards:1999bm} the spectral density at zero of the Hermitian 
Wilson Dirac operator,
\begin{equation}
 H = \gamma_5 (D-1-s),
\end{equation}
was found to be non-vanishing (but exponentially decreasing as a 
function of $\beta$). Using $A^\dagger A=H^2$, the claimed behaviour
of spectrum of $A^\dagger A$ immediately follows. 

\begin{figure}[htb!]
\vspace{3mm}
\begin{center}
\begin{minipage}{155mm}
\resizebox{75mm}{!}{
\includegraphics{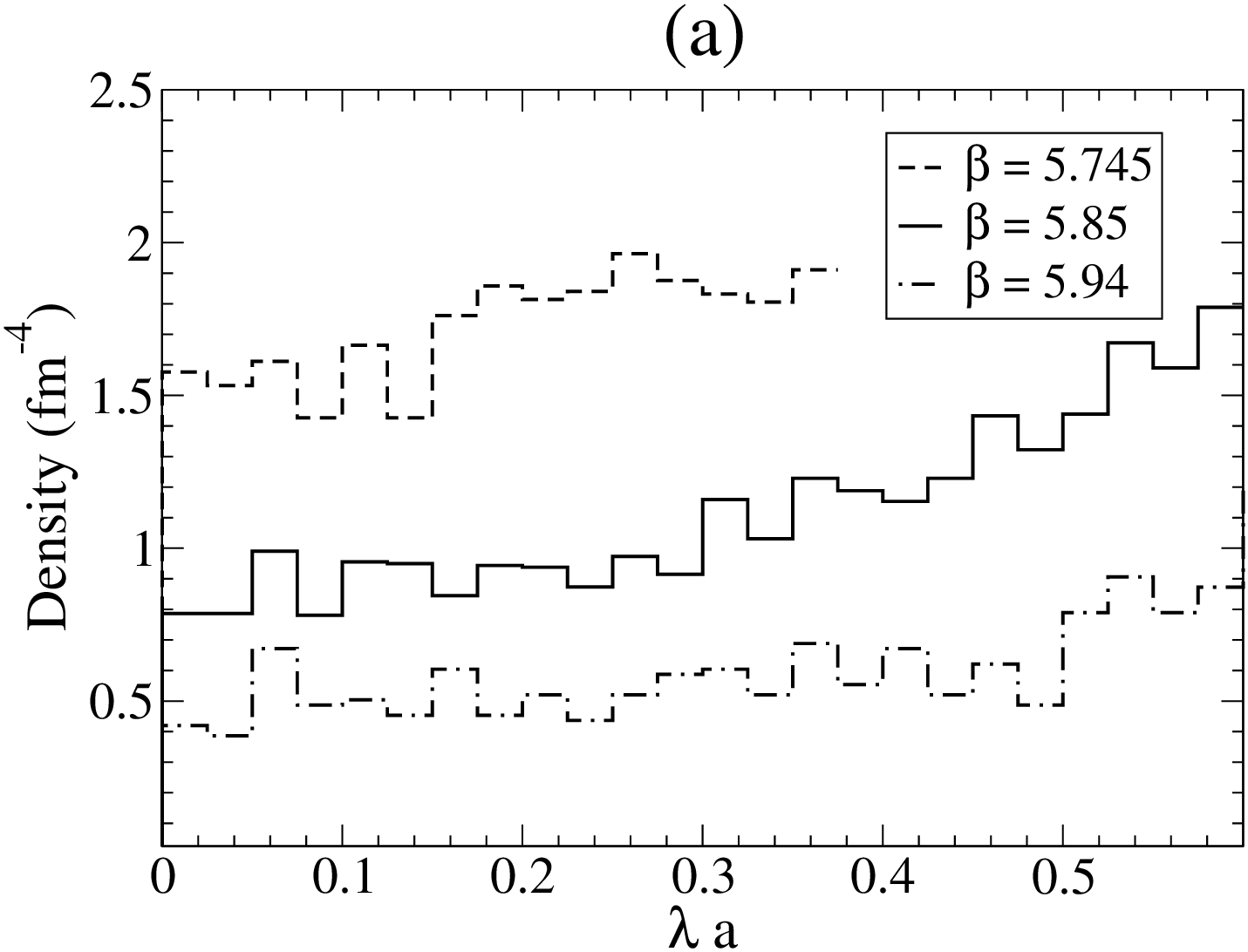}} \hfill
\resizebox{75mm}{!}{
\includegraphics{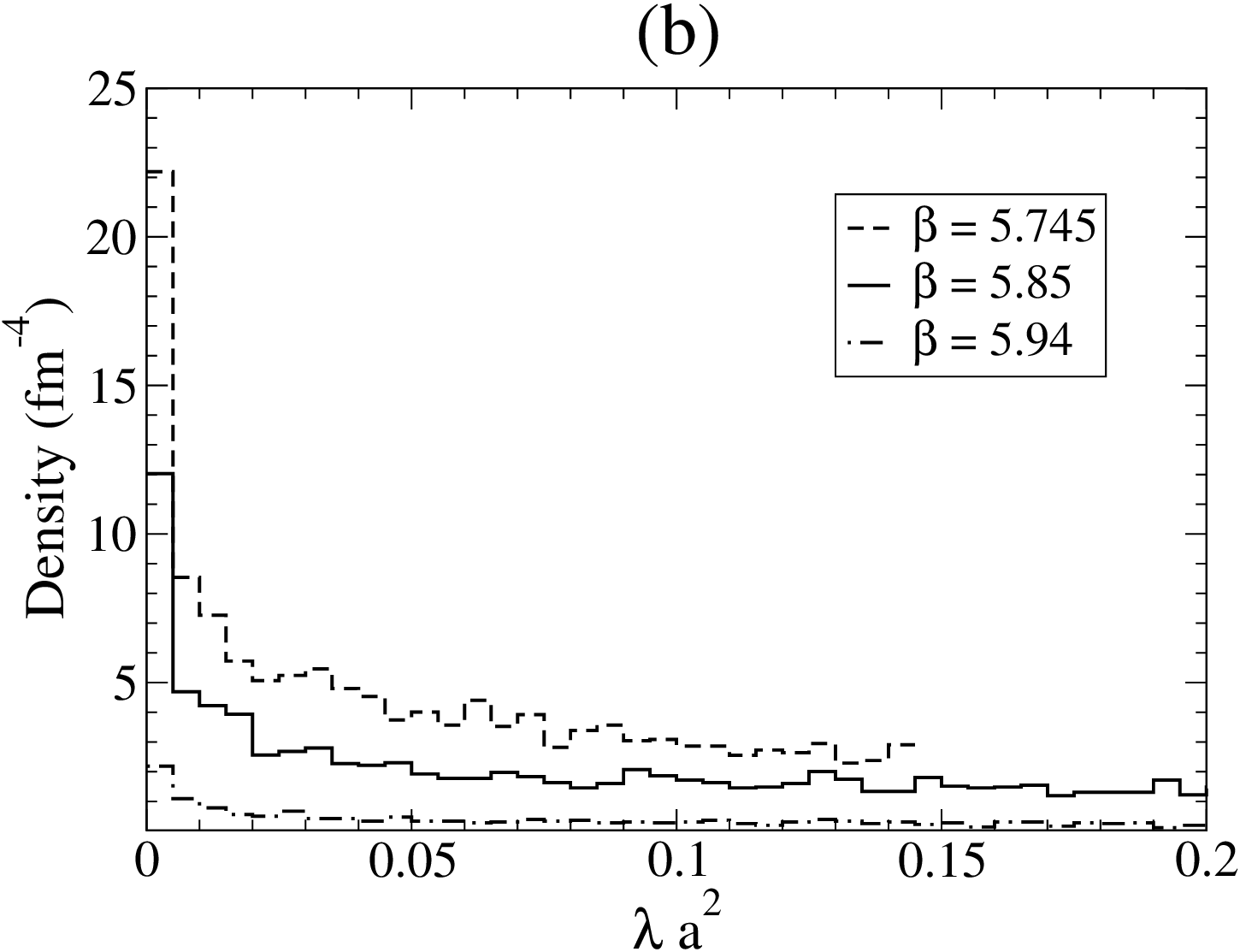}}
\vspace{-0.0cm}\caption{\label{fig:a_dens} The spectral
density of $(A^\dagger A)^{1/2}$ (a), and of $A^\dagger A$ (b),
on the APE10 gauge backgrounds of the three fixed physical volume 
ensembles of Table \ref{tab:pars}.}
\end{minipage}
\end{center}
\end{figure}

We verified this behaviour for the spectrum 
of the clover $A^\dagger A$ on APE10 quenched gauge backgrounds.
In Fig.\ \ref{fig:a_dens} we plotted the numerically obtained spectral 
density of both of $(A^\dagger A)^{1/2}$ (a), and of $A^\dagger A$ (b)
on the three fixed physical volume gauge ensembles listed in 
Table \ref{tab:pars}. The spectral density of $(A^\dagger A)^{1/2}$ 
at zero appears to be non-vanishing (Fig.\ \ref{fig:a_dens}a), and the 
expected singularity of the spectral density of $A^\dagger A$
shows up as a spike at zero (Fig.\ \ref{fig:a_dens}b). We also verified 
that this qualitative behaviour persists if less or no smearing is
applied to the gauge field and the clover term is switched off.
In the latter case, however, the numerical value of the spectral 
density of $(A^\dagger A)^{1/2}$ is almost an order of magnitude
higher.

The spectral density of $(A^\dagger A)^{1/2}$  at zero can be quite accurately 
determined as the slope of a linear fit to the integrated spectral density  
\begin{equation}
 I(\lambda a) = \int_0^{\lambda a} \rho(x) dx
\end{equation}
for small values of $\lambda$. The obtained values are listed in 
the last column of Table \ref{tab:pars}\footnote{Notice that
the spectral density was defined as the number of eigenvalues 
per fm$^4$ in terms of the dimensionless eigenvalues $\lambda a$,
therefore its unit is fm$^{-4}$.}. 
Although the $\beta$ range is a bit narrow,
this behaviour is consistent with an exponentially falling
spectral density that has been already reported for the thin link
Wilson operator over a much wider $\beta$ range \cite{Edwards:1999bm}.
Since our volumes are not particularly big, we also checked that the
spectral density obtained on the one available larger volume is 
consistent with that in the corresponding smaller volume at the same 
coupling.

Given the fact that even on these locally very smooth smeared gauge fields,
on which the average plaquette is at least 2.97, a nonzero spectral density 
of $(A^\dagger A)^{1/2}$ persists, it is very unlikely that this picture 
will qualitatively change with any reasonable gauge action or fermion 
kernel in the overlap. (Of course the numerical value of the spectral 
density can strongly depend on these details.) 
The good news is, however that the spectral density
decreases with the gauge coupling and as Fig.\ 
\ref{fig:ovrange} shows, even at moderately large couplings, these
small modes do not seem to have an important impact on the locality
of the overlap.

\begin{figure}[htb!]
\vspace{3mm}
\begin{center}
\begin{minipage}{110mm}
\resizebox{\textwidth}{!}{
\includegraphics{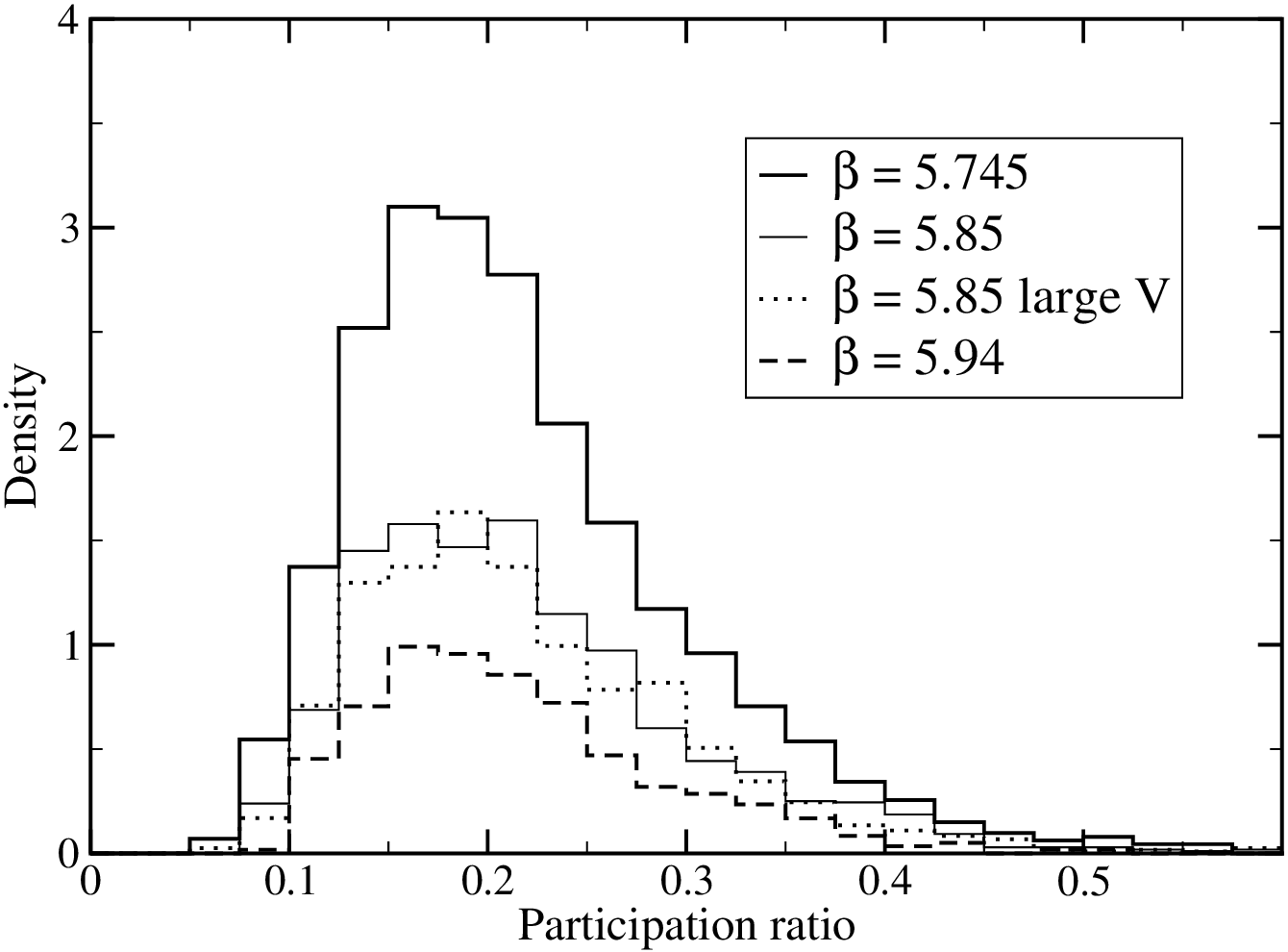}}
\vspace{-0.5cm}\caption{\label{fig:loc} The distribution of
participation ratios of low modes $(\lambda a \leq 0.1)$ of 
the APE10 FCO $A^\dagger A$ on the four gauge ensembles listed 
in Table \ref{tab:pars}.}
\end{minipage}
\end{center}
\end{figure}

To understand at least qualitatively why this is so, we can look
at the localisation of the small eigenmodes of $A^\dagger A$.
In Fig.\ \ref{fig:loc} we plotted the participation ratio distribution
of low modes of $A^\dagger A$ for the four ensembles listed in Table
\ref{tab:pars}. The participation ratio is defined as
\begin{equation}
   P = \frac{\sum_i [\psi(i)^\dagger \psi(i)]^4}
            {\left( \sum_i [\psi(i)^\dagger \psi(i)]^2 \right)^2},
\end{equation}
where $\psi$ is the wave function and the summation is over the whole
lattice. $P$ is a measure of the extension of the wave function; it can
vary between $1/V$ (if $\psi^\dagger\psi$ is homogeneously distributed over 
the whole volume) and 1 (if it is localized on a single site). The
important points that can be inferred from Fig.\ \ref{fig:loc} are that
small modes of $A^\dagger A$ are rather localized and their
localisation (in lattice units) appears to be independent of 
$\beta$ as well as of the physical volume.
This suggests that these eigenmodes correspond to independent
gauge defects that have a fixed size of the order of the cut-off.
Moreover, as we have seen, their physical density is decreasing
at weaker coupling, therefore they are incapable of propagating quarks
to large distances and are very likely to be harmless lattice
artifacts that do not influence the locality of the overlap in the
continuum limit.

Apart from these theoretical considerations, there is also
a practical reason to prefer the FCO action to the WO. We 
found that, depending on the extent and type of smearing used,
the computation of the FCO is a factor 2-5 cheaper than that
of the WO. The reason for this is that smearing considerably
raises the point where the spectrum of $A^\dagger A$ becomes
dense. In our simulation we projected out the 8 lowest eigenmodes
of $A^\dagger A$, treated them exactly and approximated the
inverse square root of the remainder with Chebyshev polynomials.
The degree of the polynomial needed for a given precision
is proportional to $\lambda_8^{-1/2}$, where $\lambda_8$ is
the 8th smallest eigenvalue of $A^\dagger A$. We compared 
$\langle \lambda_8 \rangle$ for different types of smearing
on the $\beta=5.94$ ensemble of Table \ref{tab:pars}. As
a reference we used the same quantity for the WO (no smearing,
$c\msub{sw}=0.0$) at $s=0.4$. The factors of reduction
in computation time were the following: 2.2 for APE2, 3.3 for
APE4, 4.7 for APE10, all with a smearing factor $c=0.45$.  
We also compared these to one level of so called ``HYP'' smearing,
a slightly more complicated smearing that stays within 
a hypercube \cite{Hasenfratz:2001hp}. Here we found an improvement
factor of 3.7 compared to the WO.  Also in other respects 
(locality of the overlap, density of
small modes of $(A^\dagger A)^{-1/2}$) HYP smearing seems to have 
similar or slightly better properties than APE4 smearing,
but it is strictly localized within a hypercube.
In this respect HYP smearing probably represents the optimal 
choice so far for building a FCO action.

\section{Topology}
    \label{sec:T}
\subsection{Zero mode wave functions}

It has been observed that Wilson overlap quark zero mode wave 
functions corresponding to smooth instantons are considerably
more extended than their continuum counterparts 
\cite{Gattringer:2001cf}. While this effect is expected to decrease
for larger instantons, it can still be important in
current lattice simulations where typical instantons
are on the scale of a few lattice spacings. 

\begin{figure}[htb!]
\vspace{3mm}
\begin{center}
\begin{minipage}{120mm}
\resizebox{\textwidth}{!}{
\includegraphics{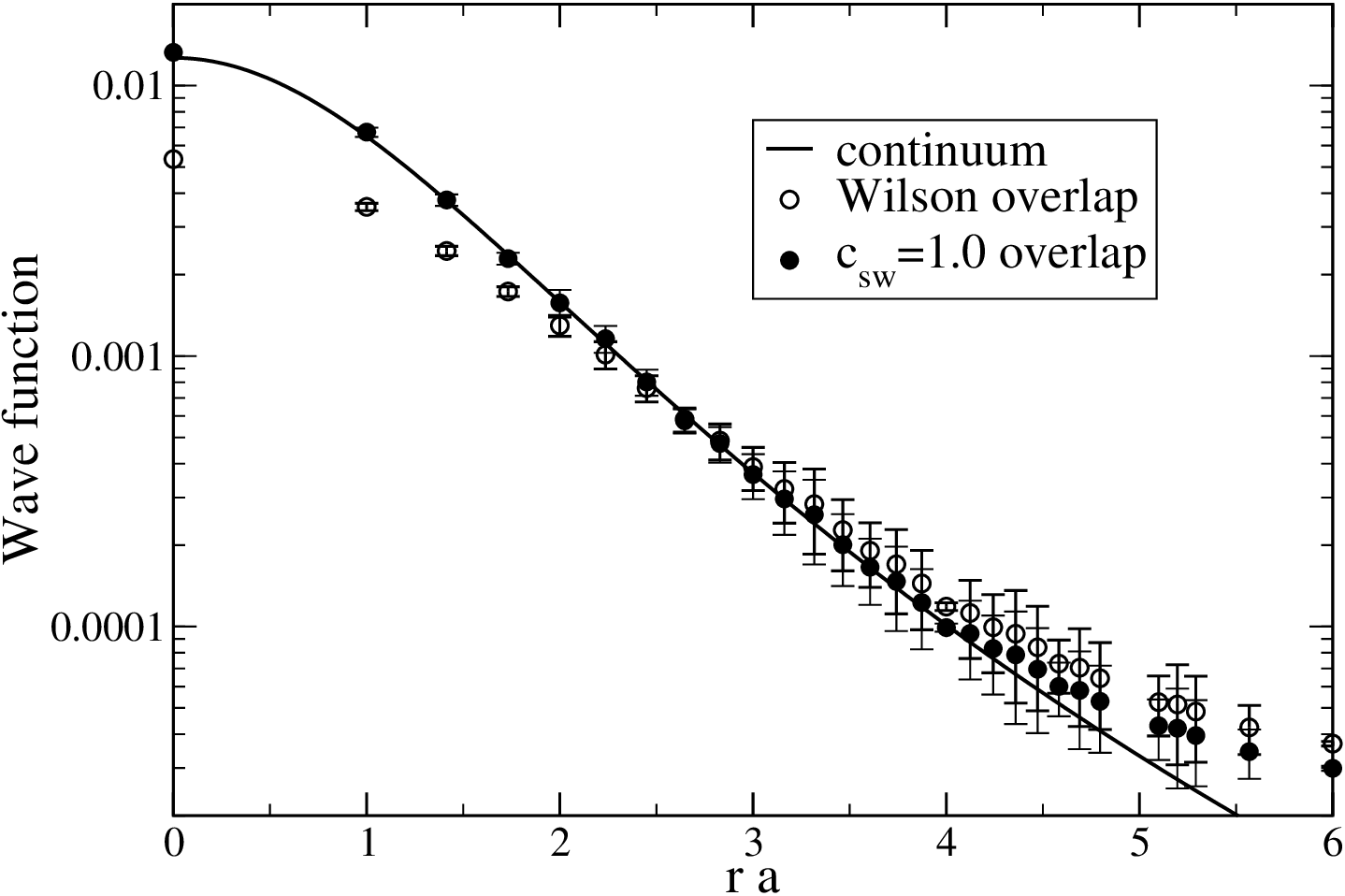}}
\vspace{-0.5cm}\caption{\label{fig:id2.0} The zero mode wave 
function $\psi^\dagger \psi(x)$ as a function of the distance
from the centre of a size $\rho/a =2.0$ instanton. Filled
symbols represent the $c\msub{sw}=1.0$ clover overlap, open symbols
correspond to the Wilson overlap and the continuous line shows
the analytical continuum expression.}
\end{minipage}
\end{center}
\end{figure}

To see how our FCO action fares in this respect, we compared
the FCO, the WO and the exact analytical continuum zero mode
wave function on discretised lattice instantons of various sizes.
Since the instanton backgrounds were locally smooth, we did not
apply any smearing in this study, that would not have changed
our results appreciably. Both for the FCO and the WO operator
time antiperiodic and space periodic boundary conditions were chosen.

Fig.\ \ref{fig:id2.0} shows a typical result for and instanton size
of $\rho/a = 2.0$. The values of $\psi^\dagger \psi(x)$ have been
averaged over all the lattice points at a distance $|x|=r$ from
the centre of the instanton. Error bars represent the standard
deviation of $\psi^\dagger \psi(x)$ at a given distance
from the centre. It is non-zero partly because
the lattice instanton was constructed a bit off-site and also 
because different directions are not equivalent; spherical symmetry 
is broken by the hypercubic geometry of the finite box.

We can confirm that the 
WO zero modes are much more extended
than the continuum zero modes, as claimed in Ref.\ 
\cite{Gattringer:2001cf}. On the other hand the FCO 
zero modes follow very closely the continuum wave functions
even for instanton sizes as small as $a \leq \rho \leq 2a$.

\subsection{Index theorem}

The Atiyah-Singer index theorem relates the topological charge
$Q$ of a smooth continuum gauge field to the number of zero
modes of the Dirac operator as
\begin{equation}
  Q = N_- - N_+,
\end{equation}
where $N_+$ and $N_-$ are the number of positive and negative 
chirality zero modes \cite{Atiyah:rm}.
Since the overlap has exact chiral zero modes, it is interesting
to ask whether their number with a certain gauge 
field definition of the charge will satisfy the index theorem. 

For the gauge field definition of $Q$ we use the ``Boulder charge''
which is constructed from two loops in two representations
evaluated on smeared gauge links. Details of the construction and its
motivation can be found in Ref.\ \cite{Hasenfratz:1998qk}.
The smearing we chose to use here was 8 steps of APE smearing with
$c=0.45$.

At any finite gauge coupling an exact index theorem cannot be
expected to hold. The best one can hope for is that the index
theorem be more and more precisely satisfied as the continuum
limit is approached. The reason for this is that small topological 
objects on the scale of the cut-off can occur with a finite
probability. Whether a given charge operator will identify
such an object or not, can depend on the non-universal
details of the operator in question.

It might be possible to detect the occurrence of such an object 
both with the gauge field charge and the
Dirac operator. A topological object on the cut-off
scale corresponds to a configuration close to the boundary
between two charge sectors. Since the gauge field charge is
a continuous function of the field, such an 
object should be signalled by a gauge field charge that is 
not close to any integer value. In terms of the overlap,
transitions between different topological sectors are
singular points. On the other hand, the operator $A^\dagger A$ 
depends continuously on the gauge field and it has a zero eigenvalue
whenever the number of zero modes of the overlap 
change. Therefore, if the gauge field is in the vicinity of 
a transition between two charge sectors, it is signalled by a
small eigenvalue of $A^\dagger A$. 

Based on this intuitive picture, we shall loosely say that a 
small topological object is detected by a certain charge 
definition if according to that charge definition the gauge
field is ``close'' to the boundary between different charge 
sectors. Two topological charge definitions, be they gauge field 
or fermionic, can be consistent in the continuum limit only if 
they have the following two properties. 

\begin{enumerate}
\item[(A)] The density, per unit physical volume, of small,
ambiguous topological objects must go to zero in the continuum
limit for both definitions of the charge.
\item[(B)] If neither charge operator detects a small topological
object, they have to give the same charge value.
\end{enumerate}

\begin{table}[htb!]
\begin{center}
\vspace{3mm}
\begin{tabular}{|l||r|r|r|r|r|r|} \hline
 $\beta$ & size &  $a$ (fm) & $a^2 \lambda\msub{min} \leq 0.5$ & 
$a^2 \lambda\msub{min} \leq 0.2$ & $\Delta Q \geq 0.1$ &  $\Delta Q \geq 0.2$  
                                                            \\ \hline\hline
 5.745 & $8^4$  &  0.153 & 0.97  & 0.824 &  n.a. & n.a.\  \\ \hline
 5.85  & $10^4$ &  0.123 & 0.863 & 0.637 & 0.652 & 0.457  \\ \hline
 5.94  & $12^4$ &  0.104 & 0.676 & 0.431 & 0.497 & 0.335  \\ \hline
 6.033 & $14^4$ &  0.088 & 0.456 & 0.243 & 0.335 & 0.210  \\ \hline\hline
\end{tabular}
\end{center}
  \caption{The fraction of configurations ``close'' to the
boundary between charge sectors, according to different criteria.
$\lambda\msub{min}$ is the smallest eigenvalue of $A^\dagger A$,
$\Delta Q$ is the distance of the Boulder charge $Q$ from the nearest
integer. ($Q$ was not computed on the coarsest ensemble.) \label{tab:top}}
\end{table}

At first sight property (A) might seem trivially true, but
in fact there are examples when it does not hold. This is the
case e.g.\ in quenched SU(2) gauge theory with the Wilson
action and the geometric charge. The dislocation problem 
\cite{Pugh:ek}, appearing there, is essentially equivalent 
to the violation of property (A). Whether (A) is satisfied in
a particular case, depends both on the action and the charge
operator.  

In Table \ref{tab:top} we demonstrate how the probability of 
configurations with an ``ill defined'' charge changes with $\beta$. 
For the fermionic definition, we listed the fraction of 
configurations on which the smallest eigenvalue of $A^\dagger A$ 
was below 0.5 (0.2). For the gauge field definition we computed
the fraction of configurations with a charge farther than
0.2 (0.1) from the nearest integer. According to all these 
criteria the charge appears to become more well-defined
for weaker coupling. We emphasize that throughout this study
the number of smearing steps was kept fixed at 10 (and 8) for
the fermionic (and the gauge) definition of the charge. 

\begin{figure}[htb!]
\vspace{3mm}
\begin{center}
\begin{minipage}{120mm}
\resizebox{\textwidth}{!}{
\includegraphics{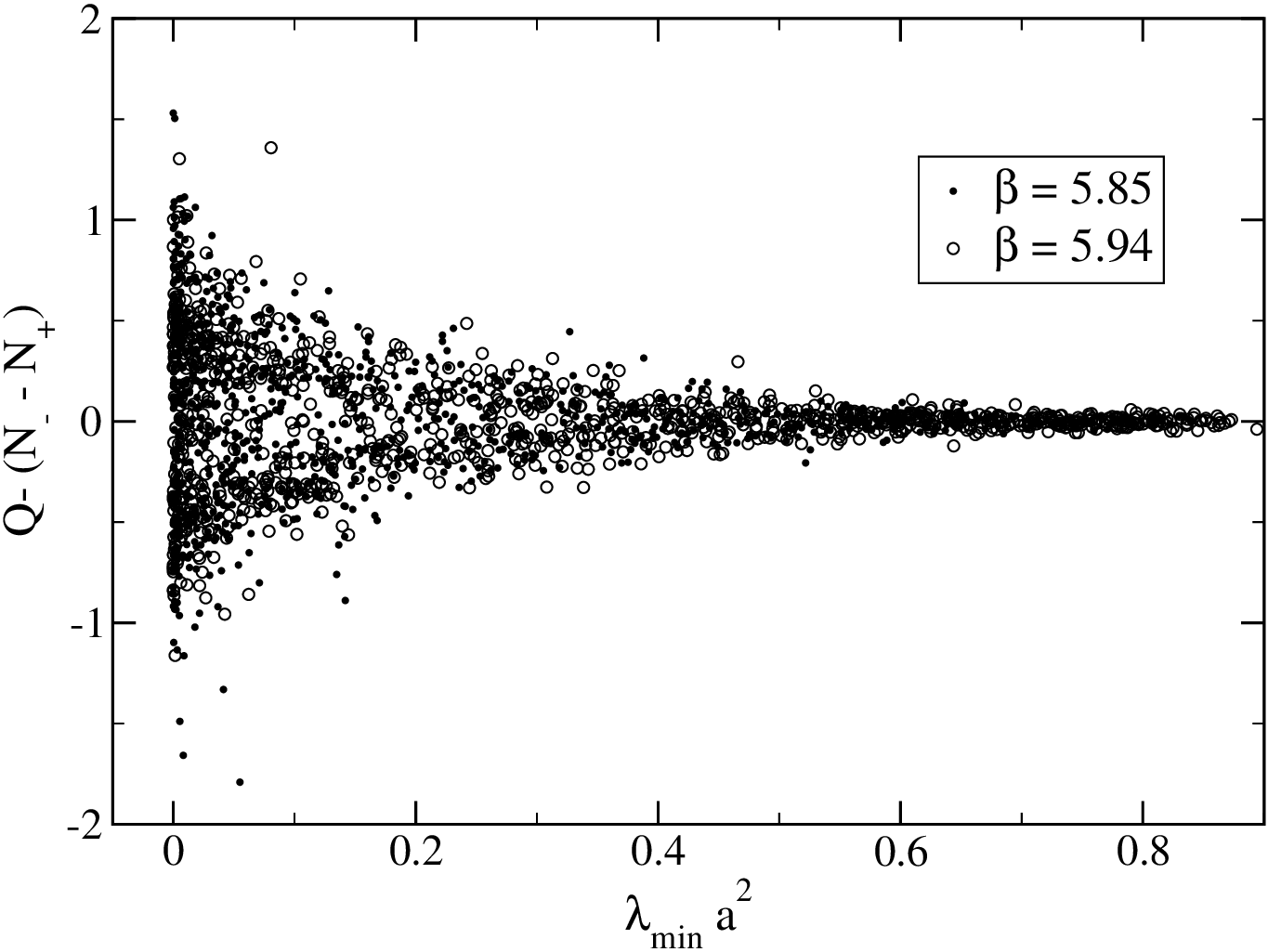}}
\vspace{-0.5cm}\caption{\label{fig:qvsl} Scatter plot of 
the difference between the APE10 FCO overlap zero mode charge and the 
gauge field charge versus the smallest eigenvalue of $A^\dagger A$. 
Two ensembles of 1000 configurations were plotted from Table \ref{tab:pars}.}
\end{minipage}
\end{center}
\end{figure}

It was also essential to keep the physical volume of
the compared configurations constant as $\beta$ was increased. 
If our intuitive picture is correct, 
and the boundary between charge sectors is associated
with localized small defects, we expect that in larger
volumes the charge becomes ill defined more frequently
(at fixed $\beta$).

Let us see whether our gauge and fermionic charge definition 
satisfy criterion (B). In Fig.\ \ref{fig:qvsl} we show
a scatter plot of the difference between the fermionic
and the gauge field charge versus the smallest eigenvalue 
of $A^\dagger A$ on 2000 gauge configurations obtained at
two values of $\beta$ on the same physical volume.
It is clear that if $a^2 \lambda\msub{min}$ is larger than
about $0.4-0.6$, the two charge definitions are rather close.
This means that (B) is also satisfied.

Interestingly enough, on these locally smooth gauge backgrounds
there is a good correlation between small eigenmodes of $A^\dagger A$
and change of topology. We would like to remark that while
the latter implies the former by continuity, the argument
does not hold the other way round. It is not guaranteed that
a small mode of $A^\dagger A$ implies that the gauge
configuration is close to the boundary between charge sectors.
In fact, on unsmeared configurations the density of small
modes is much larger and it is possible that many or even
most of these small modes are not related to topology.

\subsection{Topological susceptibility}

\begin{figure}[htb!]
\vspace{3mm}
\begin{center}
\begin{minipage}{110mm}
\resizebox{\textwidth}{!}{
\includegraphics{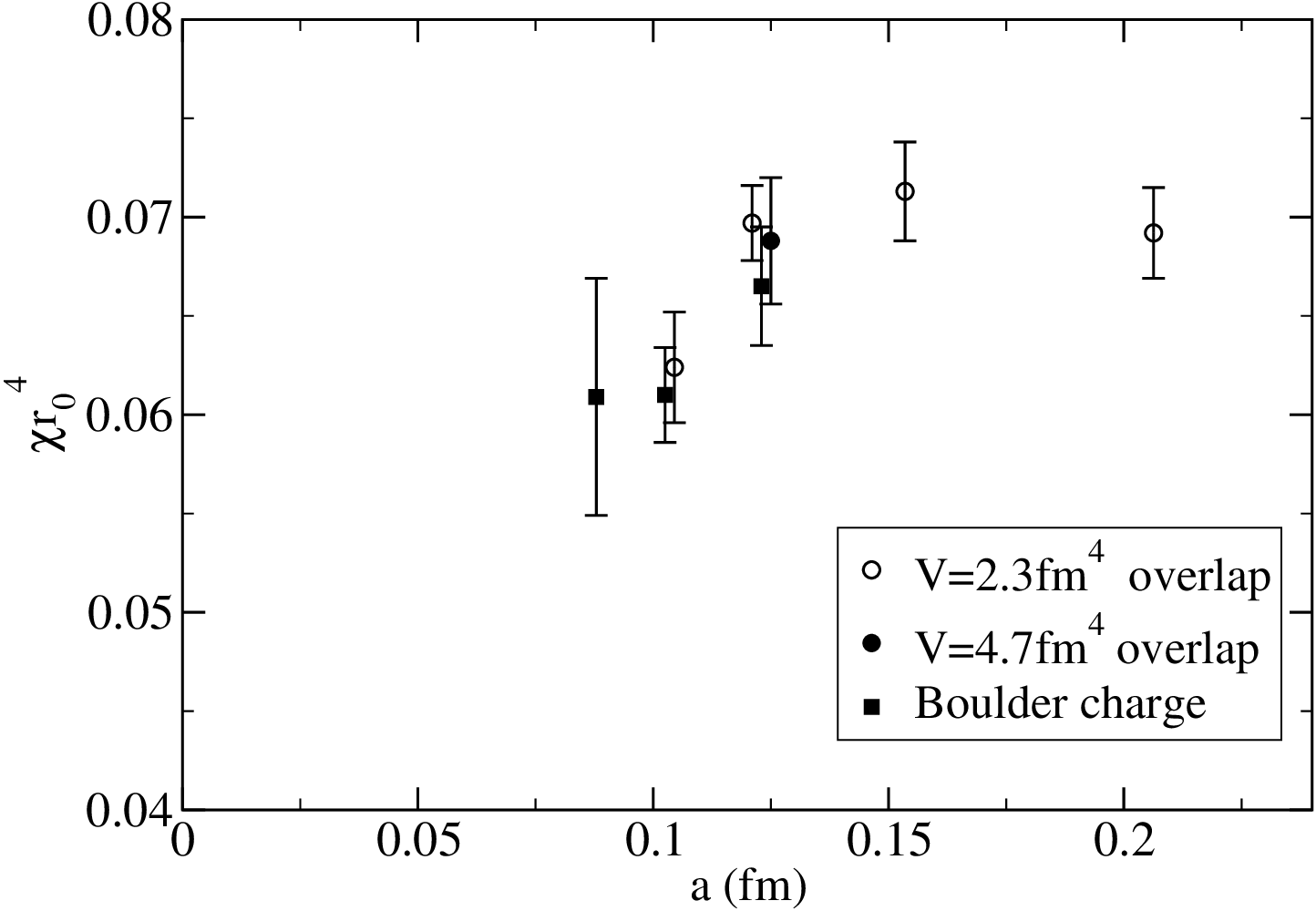}}
\vspace{-0.5cm}\caption{\label{fig:susc} The topological susceptibility
as a function of the lattice spacing. The scale was set with the Sommer
parameter $r_0=0.49$ fm.}
\end{minipage}
\end{center}
\end{figure}

We also computed the quenched zero mode susceptibility 
\begin{equation}
  \chi\msub{zm} = \frac{1}{V} \langle (N_+ - N_-)^2 \rangle,
\end{equation}
with the APE10 FCO action as well as the topological susceptibility 
\begin{equation}
  \chi\msub{top} = \frac{1}{V} \langle Q^2 \rangle,
\end{equation}
with the APE8 Boulder charge. The results are summarized in 
Fig.\ \ref{fig:susc}. In addition to the ensembles listed in
Table \ref{tab:pars}, here we also used a coarser ensemble
of 1000 $6^4$ lattices with $\beta=5.62$ and a finer one
of 200 $14^4$ lattices at $\beta=6.033$. On the two coarsest
ensembles the Boulder charge was not computed, while on
the finest one, we computed only the Boulder charge. The
physical volumes were chosen to be approximately the same,
except for the $12^4$ $\beta=5.85$ ensemble where the volume
was about two times bigger to test finite volume effects. Although
the fixed volumes $V \approx 2.3$ fm$^4$ are not particularly
big, no finite size effects can be observed. After the discussion
in the previous subsection, it does not come as a surprise
that the Boulder charge susceptibility and the overlap zero mode
susceptibility appear to be compatible whenever both are 
available.

There is no simple trend as to how the susceptibility changes
with $\beta$, most likely because our $\beta$ range at weak
couplings is rather limited. For this reason we do
not attempt a continuum extrapolation. Our results, however, at the
weakest two couplings are consistent with those of the
other available high statistics chiral fermion computation 
of the quenched susceptibility \cite{Hasenfratz:2002rp}.

\section{Conclusions}

In the present paper we studied the properties of fat link
clover overlap actions. The motivation came from the already
known good properties of fat link clover actions \cite{DeGrand:1998mn}.
It turns out that if used in the overlap construction,
much less smearing (2-4 steps of APE smearing or 1 step of
HYP smearing) is enough to guarantee a substantial improvement
in several properties. In particular, we showed here that
the localisation range of the FCO action in quenched gauge
backgrounds is independent of the gauge coupling in the range of 
Wilson $5.75 \leq \beta \leq 6.00$, suggesting that the FCO
remains local in the continuum limit.

Another advantage of the construction
is that in contrast to the Wilson overlap (WO), the parameter
$s$ does not need to be optimized, one can simply use the
tree level value $s=0.0$. The FCO also turns out to be 
slightly more local than the WO at the optimal value of
$s$. The improvement in this respect, however, is not 
nearly as significant as in the case of more complicated
fermion actions \cite{Hasenfratz:2002rp,DeGrand:2000tf,Bietenholz:2002ks}.
Smearing also improves the condition number of $A^\dagger A$
and results in a saving of a factor of 2-5 in CPU time when
evaluating the overlap. Finally, in contrast to the WO,
the FCO action correctly describes fermion zero mode wave functions 
of instantons as small as the cut-off. 

We explicitly studied the locality of the involved
smearing. We found that 2-4 steps of APE 
smearing (with a smearing coefficient of
$c=0.45$) can be safely used; the range of the smearing
is bounded with an exponential with an exponent much bigger
than 1 (in lattice units). For some purposes, as much 
as 10 smearing steps might still be acceptable. Probably the 
best compromise between locality and efficiency is 
hypercubic smearing that mixes only gauge degrees of freedom
within a hypercube \cite{Hasenfratz:2001hp}. 

We also studied the connection between small modes of
$A^\dagger A$ and the locality of the overlap. Smearing 
together with the clover term considerably reduces the
density of low modes of $A^\dagger A$ but the spectral
density still appears to be divergent at zero. This is 
a simple consequence of the fact that the Hermitian 
Dirac operator $H=\gamma_5(D-1)$ has a non-vanishing 
spectral density at zero.  Our results
show that small modes of $H$ (or $A^\dagger A$) do not 
affect the locality of the overlap in practice. The reason
for this is that the low modes are typically localized
at the scale of the cut-off, their localisation
(in lattice units) does not change as the lattice
spacing shrinks, but their physical density rapidly 
decreases at the same time. 

Low modes of $A^\dagger A$ can signal that the gauge field
configuration is ``close'' to the boundary between topological
sectors, as defined with the corresponding overlap. We also
used the Boulder charge (a field theoretic definition) to
monitor whether the gauge field is close to a transition
between charge sectors. We found that whenever the gauge
field is far away from such a transition according to
both the fermionic and the gauge field definition, the
two charge definitions are always consistent. Moreover,
the occurrence of configurations close to the transition
region becomes more rare as $\beta$ increases, while the 
physical volume stays fixed. This suggests that in the continuum
limit the FCO fermionic and Boulder field theoretic charge
satisfy the index theorem. In this connection we also 
discussed the general criteria for two charge definitions
to give consistent results in the continuum limit.
Finally we also presented a high statistics
computation of the quenched topological susceptibility 
with the fermionic definition involving the FCO action.

Fat link clover overlap actions can represent the optimal
compromise between complexity on the one hand and good 
physical properties and efficiency on the other hand. 
This might be the case also in future dynamical overlap 
simulations, especially if hybrid Monte Carlo cannot be
implemented with the overlap. 

\section*{Acknowledgments}

I would like to thank the MILC collaboration \cite{MILC}
and the creators of the ARPACK package \cite{ARPACK} for making their
code publicly available, as well as Anna Hasenfratz
for discussions and for providing the code for HYP blocking.
This work was supported by the EU's Human Potential Program under 
contract HPRN-CT-2000-00145, by Hungarian science grant OTKA-T032501,
and also partly by a Bolyai Fellowship.


\begin{thebibliography}{99}
\bibitem{MILC}
The MILC code is available at
ftp://ftp.physics.utah.edu/pub/milc/freeHEP/.
\end{thebibliography}
\end{document}